\documentclass{emulateapj}

\usepackage{natbib}
\usepackage{graphicx}
\usepackage{xcolor}
\usepackage{graphicx,epsfig}

\def\be{\begin{equation}}
\def\ee{\end{equation}}
\def\ba{\begin{eqnarray}}
\def\ea{\end{eqnarray}}

\begin{document}

\title{The Relation Between Transverse and Radial Velocity Distributions
for Observations of an Isotropic Velocity Field}

\author{Robert J. Scherrer}
\affiliation{Department of Physics and Astronomy, Vanderbilt University,
Nashville, TN  ~~37235}
\author{Abraham Loeb}
\affiliation{Harvard-Smithsonian Center for Astrophysics, 60 Garden Street,
Cambridge, MA 02138}

\begin{abstract}
We examine the case of a random isotropic velocity field, in which one of the velocity components (the ``radial" component,
with magnitude $v_z$)
can be measured easily, while measurement of the velocity perpendicular to this component (the ``transverse" component,
with magnitude $v_T$) is more difficult and requires long-time monitoring.  Particularly important
examples are the motion of galaxies at cosmological distances and the interpretation of
Gaia data on the proper motion of stars in globular clusters and dwarf galaxies. We address two questions:  what is the
probability distribution
of $v_T$ for a given $v_z$, and for what choice of $v_z$
is the expected value of $v_T$ maximized?  We show that, for a given $v_z$, the probability that $v_T$ exceeds some
value $v_0$ is $p(v_T \ge v_0 | v_z) = {p_z(\sqrt{v_0^2 + v_z^2}})/{p_z(v_z)}$, where $p_z(v_z)$ is the probability
distribution of $v_z$.  The expected value of $v_T$ is
maximized by choosing $v_z$ as large as possible whenever $\ln p_z(\sqrt{v_z})$ has a positive second
derivative, and by taking $v_z$ as small as possible when this second derivative is negative.

\end{abstract}


\maketitle

\section{Introduction}

Measuring radial velocities in astronomy through Doppler shifts in spectral lines is easy.
Measuring transverse velocities through shifts in angular position is more challenging.
Redshift measurements allowed Vesto Slipher
to determine the radial velocities of distant galaxies more than 100 years ago, while
the transverse motion of galaxies at cosmological distances
has never been measured.\footnote{The transverse velocities of the much closer LMC and SMC
have both been measured using the Hubble Space Telescope (Kallivayalil et al. 2006;
Kallivayalil, van der Marel, \& Alcock 2006).}  However, the latter situation may soon
change (Darling, Truebenbach, \& Payne 2018).

As noted previously by Sandage (1962) and Loeb (1998), precision redshift measurements taken over
a significant time span would allow for a ``real time" measurement of the evolution of the Hubble parameter;
there have been attempts to measure this effect using \ion{H}{1} 21 cm absorption line redshifts (Darling
2012).
Similarly, precision astrometry might soon allow for the measurement of galactic proper motions (and
hence, galaxy transverse velocities) in real time (Peebles et al. 2001; Nusser, Branchini, \& Davis 2012;
Quercellini 2012; Darling \& Truebenbach 2018; Darling, et al. 2018).  [The measurement of tranvserse
velocities of distant galaxies using microlensing was
considered by Grieger, Kayser, \& Refsdal (1986) and Gould (1995),
while Hamden, et al. (2010) explored the possibility of using
perspective rotation in clusters].  The possibility of making transverse velocity measurements
with Gaia is discussed in detail by Nusser et al. (2012), while Darling et al. (2018) examine the potential
for ngVLA to measure these transverse velocities. 

This leads to an obvious question:  given the opportunity to monitor a limited set of galaxies with known
radial velocities $v_z$, which of these are most likely to have the largest transverse velocities $v_T$?  (Here $v_z$
corresponds to the peculiar line-of-sight velocity, with the contribution from the Hubble expansion subtracted
off.)
This question is perhaps less relevant for an all-sky survey such as Gaia, but other instruments
such as ngVLA might monitor a limited
sample of distant galaxies.

Given
a wide dispersion in the magnitudes of the total velocity, $v$, one might naturally
assume that the galaxies with the largest radial velocities would also tend to have the largest
transverse velocities.  However, if $v$ is narrowly distributed around a single value, then
the largest radial velocities would correspond to the {\it smallest} transverse velocities.  This is easily
seen for the case where $v$ is identically the same for all of the objects in question; in this
case $v_T = \sqrt{v^2 - v_z^2}$ is maximized when $v_z=0$.

Which of these two arguments
is correct?  Both of them are relevant.  As we will see, the largest transverse velocities can
correspond to either the largest or the smallest values of the radial velocity, $v_z$,
depending on the distribution of $v_z$.

This paper addresses the following questions:  given an isotropic random velocity field, along with a
known distribution of radial velocities $v_z$, what is the corresponding distribution of transverse velocities,
$v_T$, and for what choice of $v_z$ is the expected value of $v_T$ maximized?  While this discussion
is motivated within the context of galaxy velocities, these questions are quite general, and
it seems likely that our results would be applicable to other
areas of astronomy as well, such as star clusters.
We address these questions mathematically in the next section, and briefly discuss
our results in Section 3.

\section{The distribution of transverse velocities for a given radial velocity}

Consider a set of sources with an isotropic velocity distribution.  We begin by examining the relation
between the distribution of the magnitude of the total velocity, $v$, and the distribution of the
magnitude of a single component, $v_z$, where we will usually
assume that it is the latter that can be observationally inferred.
This is a well-known problem (Feller 1971, 29-33), and the results we derive here are not new.
Note that the derivation is simplified if we take
$v_z$ to be the {\it magnitude} of a single component of the velocity, rather than the actual (positive
or negative) value of the velocity component; this will be our convention
throughout most of this paper.  As a single component of an isotropic velocity field is symmetrically distributed about
zero, it is trivial to go from the distribution of the magnitude of a single component to the distribution
of that component itself.

Following Feller (1971), the relation between $v$ and $v_z$ is
\begin{equation}
\label{vzv}
v_z = v |\cos \theta|,
\end{equation}
where $\theta$ is the angle between the line of sight and the velocity vector of the moving object.
Note that $|\cos \theta|$ is uniformly distributed between 0 and 1, and it is independent of $v$.
Consequently, the right-hand side of equation (\ref{vzv}) is the product of two independent
random variables, the first
with probability distribution function (PDF) $p(v)$, and the second with a uniform distribution.
Recall that for independent random variables $x$ and $y$, with PDFs $p_x(x)$ and $p_y(y)$, respectively,
the PDF of the product $z = xy$ is
\begin{equation}
p(z) = \int p_x\left(\frac{z}{y}\right) p_y(y) \frac{dy}{|y|}
\end{equation}
Then the PDF for $v_z$, which we will denote $p_z(v_z)$, is related to $p(v)$ through 
\begin{equation}
\label{pzp}
p_z(v_z) = \int_{v_z}^\infty p(v)\frac{dv}{v}.
\end{equation}
Taking the derivative
gives $p_z^\prime (v_z) = - p(v_z)/v_z$, where
$p_z^\prime$ is the derivative of $p_z$ with respect to its argument.
But this functional equation is valid regardless of the independent variable used
in the equation, so we can set this variable to be $v$ instead of $v_z$, which gives
us $p$ in terms of $p_z$:
\begin{equation}
\label{ppz}
p(v) = - v p_z^\prime(v).
\end{equation}

For example, suppose that the distribution of radial velocities is a Gaussian, so that the distribution
of the magnitude of the radial velocities is a one-sided
 Gaussian:
\begin{equation}
\label{Gauss1}
p_z(v_z) = \sqrt{\frac{2}{\pi}}\frac{1}{\sigma}\exp\left(-\frac{v_z^2}{2 \sigma^2}\right),~~~~~v_z \ge 0,
\end{equation}
Then equation (\ref{ppz}) gives, for the distribution of the magnitude of the total velocity,
\begin{equation}
\label{Gauss2}
p(v) = \sqrt{\frac{2}{\pi}}\frac{1}{\sigma^3} v^2 \exp\left(-\frac{v^2}{2 \sigma^2}\right).
\end{equation}
Equations (\ref{Gauss1}) and (\ref{Gauss2}) correspond, of course, to the Maxwell distribution
of velocities in a gas.

While $p(v)$ can have essentially any functional form (subject to the condition that it be a
probability distribution normalized to unity), this is not the case for $p_z(v_z)$.  For instance,
equation (\ref{ppz}) implies that $p_z^\prime(v_z) < 0$:  the distribution of $v_z$ must have its maximum
value at $v_z = 0$ and decrease monotonically with $v_z$.
An observed distribution of $v_z$ that violated this condition
would imply a deviation from an isotropic velocity field.

Equations (\ref{pzp}) and (\ref{ppz}) are restatements of previously-derived results,
but now consider the question of interest to us:
for a given observed value of $v_z$, what is the distribution of $v_T$, denoted $p(v_T|v_z)$? 
Consider first the cumulative distribution function (CDF),
which is the probability that $v_T$ is less than a given value $v_0$;
this quantity is denoted $p(v_T \le v_0 | v_z)$.  It will turn out to be easier and more
useful to work with the complementary CDF, which is $p(v_T \ge v_0|v_z)$.
Since
$v^2 = v_T^2 + v_z^2$, we can express this CDF in terms of $v$ instead of $v_T$:
\begin{equation}
\label{vT}
p(v_T \ge v_0 | v_z) = p(v \ge \sqrt{v_0^2 + v_z^2} | v_z).
\end{equation}
The right-hand side can be simplified if we have an expresson for $p(v|v_z)$, the PDF of $v$ for a given fixed
value of $v_z$.  
Equation (\ref{vzv}) implies that $p(v_z|v)$ is just a uniform distribution between 0 and $v$, given by $1/v$,
so we can
use Bayes theorem:
\begin{eqnarray}
p(v|v_z) &=& p(v_z|v) \frac{p(v)}{p_z(v_z)},\\
\label{Bayes}
&=& \frac{1}{v} \frac{p(v)}{p_z(v_z)}.
\end{eqnarray}
Integrating equation (\ref{Bayes}) over $v$ while keeping $v_z$ fixed gives us the CDF in equation (\ref{vT}),
namely
\begin{equation}
p(v_T \ge v_0 | v_z) = \int_{\sqrt{v_0^2 + v_z^2}}^\infty \frac{1}{v} \frac{p(v)}{p_z(v_z)} dv.
\end{equation}
Using equation (\ref{pzp}), we can perform the integral to give
\begin{equation}
\label{CDF}
p(v_T \ge v_0 | v_z) = \frac{p_z(\sqrt{v_0^2 + v_z^2)}}{p_z(v_z)}.
\end{equation}
Equation (\ref{CDF}) provides complete information about the distribution of $v_T$ for a given $v_z$,
and it will be the main expression we work with.  However, for completeness we will also derive
the corresponding PDF.  This is just the negative of the derivative of the right-hand side of
equation (\ref{CDF}) with respect to $v_0$, evaluated at $v_0 = v_T$.  We obtain
\begin{eqnarray}
\label{PDF1}
p(v_T|v_z) &=& - \frac{v_T}{\sqrt{v_T^2 + v_z^2}} \frac{p_z^\prime(\sqrt{v_T^2 + v_z^2})}{p_z(v_z)},\\
\label{PDF2}
&=& \frac{v_T}{v_T^2 + v_z^2}\frac{p(\sqrt{v_T^2 + v_z^2})}{p_z(v_z)},
\end{eqnarray}
where we used equation (\ref{ppz}) to go from equation (\ref{PDF1}) to equation (\ref{PDF2}).
Any of equations (\ref{CDF})-(\ref{PDF2}) provides all of the information on the distribution of transverse
velocities for an observed radial velocity, but equation (\ref{CDF}) is the simplest and most informative
of these.

As an example, consider again a Gaussian distribution of radial velocities, with the distribution
of the magnitudes of these radial velocities given by equation (\ref{Gauss1}).  Substituting this distribution
into equation (\ref{CDF}), we obtain
\begin{equation}
p(v_T \ge v_0 | v_z) = \exp(-v_0^2/2\sigma^2).
\end{equation}
Thus, for the special case of a Gaussian distribution of radial velocities, the distribution of $v_T$ is also
a Gaussian, and it is independent of the value of $v_z$.

Now we will address our second question of interest:  for what choice of $v_z$ is the expected value of
$v_T$ maximized?  From equation (\ref{CDF}), we see that there are two main possible cases.
If $p_z(\sqrt{v_0^2 + v_z^2})/p_z(v_z)$ is an increasing function of $v_z$, then the probability of observing
a value of $v_T$ greater than $v_0$ always {\it increases} with $v_z$ for any value of $v_0$.
In this case, choosing the largest value of $v_z$ will maximize the expected value of $v_T$.
Conversely,
if $p_z(\sqrt{v_0^2 + v_z^2})/p_z(v_z)$ is a decreasing function of $v_z$, then
the probability of observing $v_T$ greater than $v_0$ always decreases with $v_z$ for any value of $v_0$,
and the expected value of $v_T$ is maximized for the smallest observed value of $v_z$.  (The theoretical
best value in this case is $v_z = 0$).  These are the two possibilities discussed in
the introduction.  It is also possible for $p_z(\sqrt{v_0^2 + v_z^2})/p_z(v_z)$ to
have a local maximum or minimum, i.e., to not depend
monotonically on $v_z$, but this requires rather contrived distributions of $v_z$ and will be discussed
separately below.

We can express these conditions on $p_z(v_z)$ more simply by writing the distribution in the form
\begin{equation}
\label{fv}
p_z(v_z) = \exp[f(v_z^2)],
\end{equation}
where
the function $f$ is defined by equation (\ref{fv}), or, alternatively, by
\begin{equation}
\label{fv2}
f(v_z) = \ln p_z(\sqrt{v_z}).
\end{equation}
When $p_z(v_z)$ is written in this way, our conditions for maximizing $v_T$ become particularly transparent.
The quantity $p_z(\sqrt{v_0^2 + v_z^2})/p_z(v_z)$ is an increasing (decreasing) function of $v_z$ when 
$f(v_0^2+v_z^2) - f(v_z^2)$ is an increasing (decreasing) function of $v_z$.
Now consider the conditions on $f$ needed to make $f(v_0^2+v_z^2) - f(v_z^2)$ an increasing
function of $v_z$.  This will be the case when the derivative of this function with respect to $v_z$ is
positive, i.e., $f^\prime(v_0^2+v_z^2)-f^\prime(v_z^2) > 0$, where we have used the fact that $v_z >0$ by
definition, and the prime denotes the derivative of $f$ with respect to its argument.  It is clear that
this condition on $f^\prime$ will be satisfied as long as $f(v_z)$ is a convex function, i.e., having
positive second derivative.  Similarly, $f(v_0^2+v_z^2) - f(v_z^2)$ will be a decreasing function of $v_z$
as long as $f(v_z)$ is a concave function, i.e., with negative second derivative.
Thus, we maximize the expected value of $v_T$ by choosing $v_z$ as large as possible whenever
$f^{\prime\prime}(v_z) > 0$ and by choosing $v_z$ as small as possible whenever $f^{\prime\prime}(v_z) < 0$,
with $f$ defined by equations (\ref{fv}) or (\ref{fv2}).

Probability distributions for which $\ln p(\sqrt{x})$ is either always a concave function or always a convex
function have been examined previously in the context of signal processing (Benveniste, Goursat, \& Ruget
1980; Palmer, Kreutz-Delgado, \& Makeig 2010).  Palmer et al. introduced the terms strong sub-Gaussianity
and strong super-Gaussianity to refer to distributions for which $\ln p(\sqrt{x})$ is concave or convex,
respectively.  With this definition, $v_T$ will be maximized by choosing the largest value of $v_z$ when
$p_z(v_z)$ is strongly super-Gaussian, and by choosing the smallest value of $v_z$ when $p_z(v_z)$ is
strongly sub-Gaussian.

As an example, consider the Subbotin family of distributions, given by
\begin{equation}
\label{Subbotin}
p_z(v_z) = \frac{\beta}{\Gamma(1/\beta)}\exp(-v_z^\beta)
\end{equation} 
where $\beta$ is a positive constant and $\Gamma$ is the gamma function.\footnote{This family
of distributions, extended
from $-\infty$ to $\infty$, goes by a variety of other names, including the power exponential distribution,
the exponential power distribution, and
the generalized normal distribution.}
The corresponding expressions for the magnitude of the total velocity $v$ are (from equation \ref{ppz})
\begin{equation}
\label{Subbotinv}
p(v) = \frac{\beta^2}{\Gamma(1/\beta)}v^\beta \exp(-v^\beta).
\end{equation}

For these distributions, $f(v_z) = -v_z^{\beta/2}$ (plus an irrelevant constant), which is a convex function for $\beta < 2$
and a concave function for $\beta > 2$.  Thus, for $\beta < 2$, we maximize the expected value of
$v_T$ by choosing $v_z$ as large as possible, while for $\beta > 2$, we need to choose $v_z$ as small
as possible.  The case $\beta = 2$ corresponds to the Gaussian distribution, for which $f^{\prime \prime}(v_z) =
0$, so $p_z(\sqrt{v_0^2 + v_z^2})/p_z(v_z)$ is constant and $p(v_T \ge v_0 | v_z)$ is independent
of $v_z$, as have have already noted.

\begin{figure}[tbh]
\centerline{\epsfxsize=4truein\epsffile{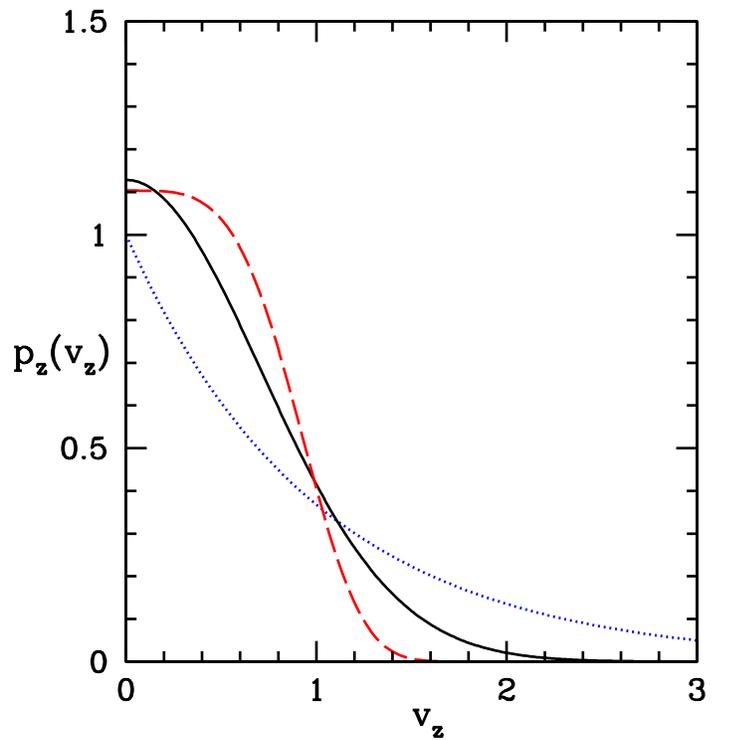}}
\caption{The Subbotin probability distribution (equation \ref{Subbotin}) for $p_z(v_z)$ as a function of $v_z$ for
$\beta = 1$ (blue, dotted), $\beta = 2$ (solid, black) and $\beta = 4$ (dashed, red).  The expected
value of $v_T$ is maximized for the largest observed value of $v_z$ for $\beta = 1$ and for the smallest
observed value of $v_z$ for $\beta = 4$.}
\end{figure}

We illustrate $p_z(v_z)$ and $p(v)$ for the Subbotin distribution in Figs. 1 and 2,
respectively, for the cases $\beta = 1$ (exponential),
$\beta = 2$ (Gaussian) and $\beta = 4$.  The form of these functions
agrees with the intuitive argument outlined
in Section 1.  The $\beta=1$ distribution for $p(v)$ has a larger tail than the Gaussian
at large $v$ and
corresponds to a case for which $v_T$ is maximized at large $v_z$.  On the other hand, the $\beta = 4$
distribution for $v$
is more sharply peaked at a single value of $v$ than is the Gaussian, and it corresponds to the case
where the expected value of $v_T$ is maximized at small $v_z$.  Indeed, in the limit $\beta \rightarrow
\infty$, the distribution of $v$ approaches a delta function in $v$, corresponding to the case discussed
in Section 1 where
$v$ is identically the same for all objects in the sample.

\begin{figure}[tbh]
\centerline{\epsfxsize=4truein\epsffile{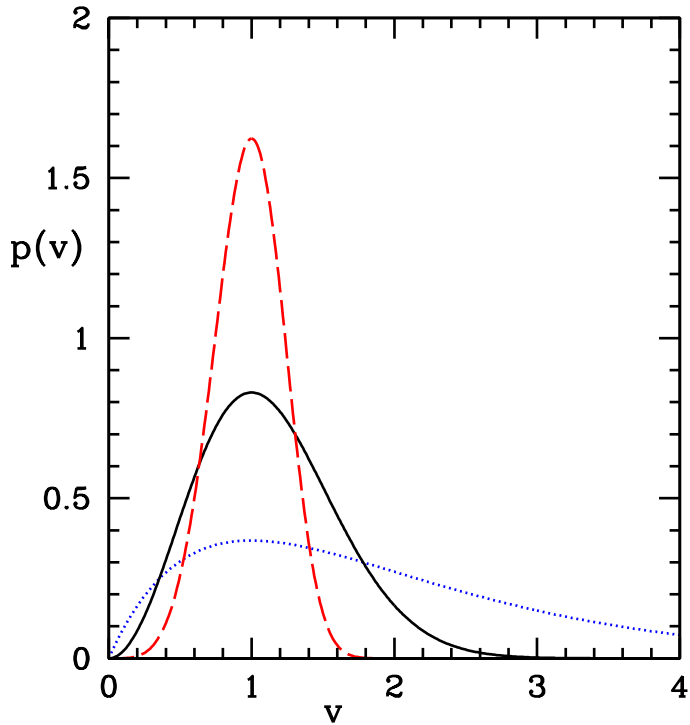}}
\caption{As Fig. 1, for the distribution of the magnitude of the total velocity, $v$ (equation
\ref{Subbotinv}) for
$\beta = 1$ (blue, dotted), $\beta = 2$ (solid, black) and $\beta = 4$ (dashed, red). The expected
value of $v_T$ is maximized for the largest observed value of $v_z$ for $\beta = 1$ and for the smallest
observed value of $v_z$ for $\beta = 4$.}
\end{figure}

Most simple monotically decreasing distributions for $p_z(v_z)$ correspond to a form for $p_z(\sqrt{v_0^2 +
v_z^2})/p_z(v_z)$ that either increases or decreases monotonically, thereby maximizing the expected
$v_T$ for the largest or smallest values of $v_z$, respectively.
These include, for example, the one-sided Gaussian distribution, the uniform distribution,
the exponential distribution, the Gamma distribution with shape parameter $< 1$, and the one-sided Cauchy
distribution.

However, these are not
the only possibilities.  Consider, for example, these two distributions for $v_z$:
\begin{equation}
\label{hybrid1}
p_z(v_z) = 1.716~\exp(-v_z - v_z^4),
\end{equation}
and
\begin{equation}
\label{hybrid2}
p_z(v_z) = 0.723~\exp\left(-\frac{v_z^4}{1+v_z^3}\right),
\end{equation} 
where the normalization constants are determined numerically.  Each of these distributions is designed to mimic
the behavior of the $\beta = 1$ and $\beta=4$ Subbotin distributions in the appropriate limits:
distribution (\ref{hybrid1}) goes to $\beta=1$ at small $v_z$ and $\beta=4$ at large $v_z$, while
distribution (\ref{hybrid2}) does the opposite. These distributions correspond to
total velocity distributions of the form
\begin{equation}
\label{hybrid1v}
p(v) = 1.716~(v+4v^4)\exp(-v - v^4),
\end{equation}
and
\begin{equation}
\label{hybrid2v}
p(v) = 0.723~\frac{v^4(v^3+4)}{(1+v^3)^2}\exp\left(-\frac{v^4}{1+v^3}\right),
\end{equation} 
respectively.
While both of these distributions are exceptionally contrived, there is nothing pathological about their
form, as can be see in Figs. 3 and 4.
\begin{figure}[tbh]
\centerline{\epsfxsize=4truein\epsffile{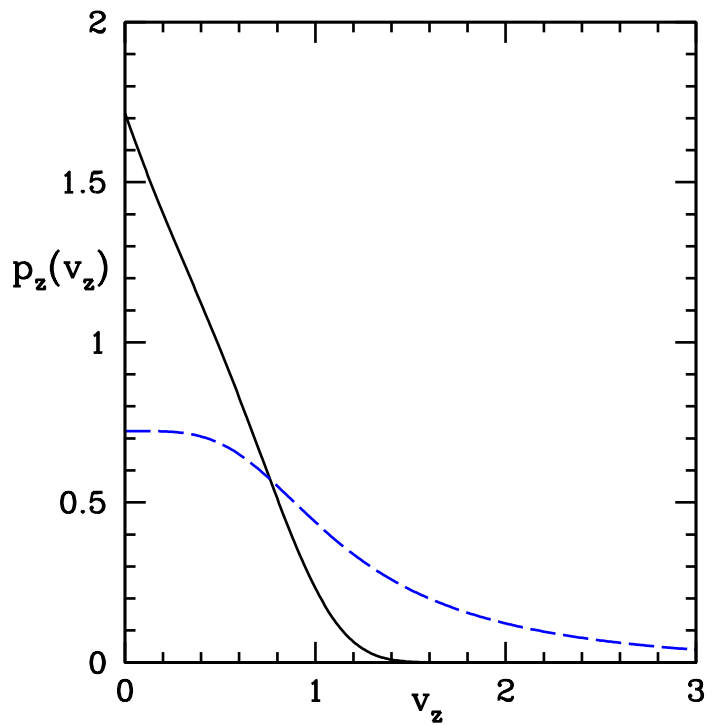}}
\caption{The distributions of the magnitude of a single component of the velocity, $v_z$ given by
equations (\ref{hybrid1}) (solid, black) and (\ref{hybrid2}) (blue, dashed).}
\end{figure}
\begin{figure}[tbh]
\centerline{\epsfxsize=4truein\epsffile{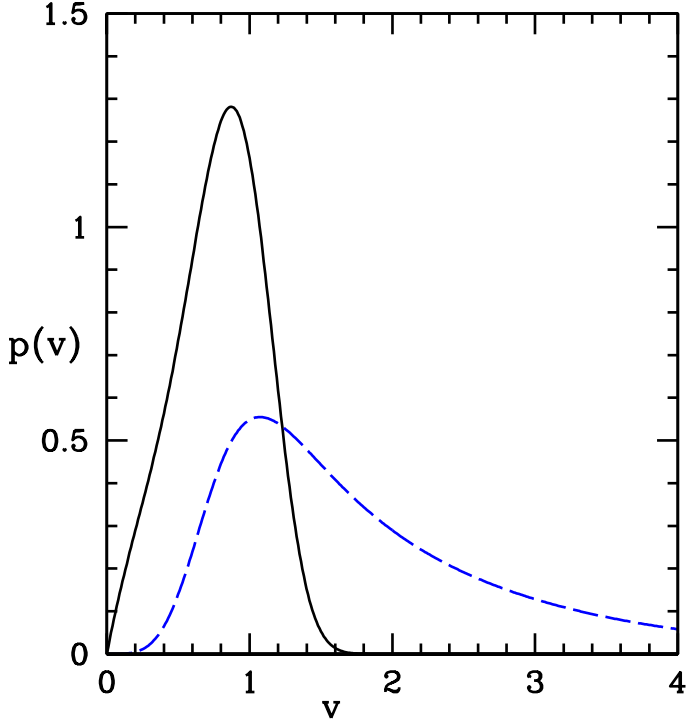}}
\caption{The distribution of the magnitude of the total velocity, $v$, given by
equations (\ref{hybrid1v}) (solid, black) and (\ref{hybrid2v}) (blue, dashed).}
\end{figure}

Now consider the behavior of $p_z(\sqrt{v_0^2 + v_z^2})/p_z(v_z)$ for these two distributions.
For the distribution given by equation (\ref{hybrid1}), our function $f(v_z)$ defined by
equations (\ref{fv}) or (\ref{fv2}) is
\begin{equation}
f(v_z) = - v_z^{1/2} - v_z^2,
\end{equation}
which has positive second derivative at small $v_z$ and negative second derivative at large $v_z$.  Thus,
the probability $p(v_T \ge v_0 | v_z) = {p_z(\sqrt{v_0^2 + v_z^2)}}/{p_z(v_z)}$ increases with $v_z$ at small $v_z$, reaches a maximum
value, and then decreases with $v_z$ at large $v_z$.  Furthermore, the value of $v_z$ at which
$p(v_T \ge v_0 | v_z)$ attains its maximum value is itself a function of $v_0$.  Thus, while
one can determine a single optimal value of $v_z$ for which the probability that $v_T$ exceeds $v_0$ is
maximized, this value of $v_z$ will now depend on $v_0$.

Conversely, for the distribution defined by equation (\ref{hybrid2}), the function $f(v_z)$ is
\begin{equation}
f(v_z) = - \frac{v_z^2}{1+ v_z^{3/2}},
\end{equation}
which has negative second derivative at small $v_z$ and positive second derivative at large $v_z$.
Consequently, $p(v_T \ge v_0 | v_z) = {p_z(\sqrt{v_0^2 + v_z^2)}}/{p_z(v_z)}$ increases as we take
either $v_z \rightarrow 0$ or $v_z \rightarrow \infty$.  Thus, we can maximize $v_T$ for a given $v_z$ by
choosing $v_z$ either as small as possible or as large as possible.

The appearance of the distributions in Fig. 1 suggests an alternate way to determine the optimal
value of $v_z$ that maximizes the expected value of $v_T$.  The Subbotin distribution with $\beta = 1$ has
positive kurtosis, while $\beta = 4$ has negative kurtosis, where we define the kurtosis
as
\begin{equation}
\kappa  = \frac{\langle v_z^4 \rangle}{\langle v_z^2\rangle^2} - 3,
\end{equation}
so that $\kappa = 0$ for the Gaussian distribution.
For our discussion of kurtosis (only) we will take $v_z$ to be the actual $z$ component of the velocity, rather
than its magnitude, so that $\langle v_z \rangle$ = 0.  Then the Subbotin distribution (equation
\ref{Subbotin}) extended to negative values of $v_z$ has kurtosis
\begin{equation}
\kappa = \frac{\Gamma(5/\beta) \Gamma(1/\beta)}{\Gamma(3/\beta)^2} - 3,
\end{equation}
which is indeed positive for $\beta < 2$ and negative for $\beta > 2$.

It might appear that kurtosis can provide a simpler criterion for the value of $v_z$ that maximizes
the expected $v_T$:  positive kurtosis distributions for $p_z(v_z)$ (which have larger tails than
a Gaussian) indicate that the largest value of $v_z$ should be chosen, while negative kurtosis
distributions point toward the smallest value of $v_z$.  However, this argument is only partially correct.
Palmer et al. (2010) show that, in fact, all strong sub-Gaussian distributions have negative kurtosis,
and all strong super-Gaussian distributions have positive kurtosis.  However, the converse
is not true.  This is obvious, since the distributions in equations (\ref{hybrid1}) and (\ref{hybrid2})
have negative and positive values of $\kappa$, respectively, and yet the first maximizes $v_T$ at a fixed
value of $v_z$, while the second maximizes $v_T$ at either large or small values of $v_z$.  Thus, while
kurtosis can provide a useful guide, the rigorously correct procedure to maximize
the expected value of $v_T$ is to maximize the right-hand side of equation
(\ref{CDF}).

\section{Discussion and Conclusions}

We have derived an expression for the distribution of the transverse velocity, $v_T$, for a given fixed value
of the radial velocity, $v_z$, valid for any isotropic velocity distribution (or indeed, for any isotropic vector field)
in equation (\ref{CDF}).  Our results indicate that the expected value for $v_T$ can be maximized by choosing the largest
possible value of $v_z$ if $v_z$ has a strongly super-Gaussian distribution, and for the smallest possible value of $v_z$
if the distribution is strongly sub-Gaussian, where these terms are defined in the previous section.

We now circle back to the question which originally motivated this investigation:  what about the peculiar velocity field of
galaxies?  While current observations are beginning to probe
this distribution (e.g., Tully, et al. 2013; Springob, et al. 2014; Tully, Courtois, \& Sorce 2016), the data are still
too noisy to provide a precise estimate of $p_z(v_z)$. The uncertainties in the
measured peculiar velocities are typically of order the velocities themselves at cosmological distances
(Watkins \& Feldman 2015). However, this problem can be mitigated by binning the velocity data.
Using the catalog of Tully, et al. (2013), Sorce (2015) derived a
bias-corrected distribution for $v_z$ which is consistent with a Gaussian distribution.  This
is precisely the unique distribution for which the value of $v_T$ is insensitive to the value of $v_z$.
It is also consistent with the theoretical model of
Sheth and Diaferio (2001), which predicts a form for $p_z(v_z)$ that looks Gaussian
at small $v_z$.  However, their model also predicts an exponential distribution for $p_z(v_z)$ at large $v_z$.
For the exponential distribution, we
expect that $v_T$ will be largest when $v_z$ is maximized.  This suggests that if one were monitoring
a limited set of galaxies over a long time span, efforts should be concentrated on those with the largest radial pecular
velocities.

Future data sets to which these results might be applied include measurements of radial peculiar velocities from
distance calibrators such as SN Ia (Riess 1999) or gravitational wave sources (Chen, Fishbach, \& Holz 2018) or
from the kinetic Sunyaev-Zel'dovich effect (Akrami, et al. 2018).  The
derivations presented here can also be used as a constraint on
models of the peculiar velocity field of galaxies in the
standard $\Lambda$CDM cosmology, such as those in Sheth \& Diaferio (2001).  In addition,
our derivations can be applied to new astrometric data from the Gaia satellite on the proper motion of
stars in globular clusters and dwarf galaxies (Helmi, et al. 2018) in an attempt to constrain their
mass distribution (Milone, et al. 2018) or rotation (Bianchini, et al. 2018) as well as the possible
existence of an intermediate black hole at their center (e.g., Kiziltan, et al. 2017).

\section*{Acknowledgments}
A.L. was supported in part by the Black Hole Initiative, which is funded by a grant from the John Templeton Foundation.
R.J.S. was supported in part by the Department of Energy
(DE-SC0019207).


\begin{references}

\reference{Akrami}
Akrami, Y., et al., 2018, arXiv:1807.06205

\reference{Benveniste}
Benveniste, A., Goursat M., Ruget, G., 1980,
IEEE Transactions on Automatic Control, 25, 385

\reference{Bianchini}
Bianchini, P., et al., 2018, \mnras, 481, 2125

\reference{Chen}
Chen, H.-Y., Fishbach, M., Holz, D.E., 2018, Nature, 562, 545 

\reference{Darlind0}
Darling, J., 2012, \apj, 761, L26

\reference{Darling1}Darling, J., Truebenbach, A.E., 2018,
\apj, 864, 37

\reference{Darling2}Darling, J., Truebenbach, A., Paine, J., 2018,
arXiv:1807.06670.

\reference{Feller}Feller, W., 1971, An Introduction to Probability Theory and Its Applications, Vol. II,
New York, NY

\reference{Grieger}
Greiger, B., Kayser, R., Refsdal, S., 1986,
Nature, 324, 126

\reference{Gould}
Gould, A., 1995, \apj, 444, 556

\reference{Hamden}
Hamden, E.T., Simpson, C.M., Johnston, K.V., Lee D.M., 2010,
\apj, 716, L205

\reference{Helmi}
Helmi, A., et al. (Gaia Collaboration), 2018,
\aap, 616, A12

\reference{Kallivayalil1}
Kallivayalil, N., van der Marel, R.P., Alcock, C.,
Axelrod, T., Cook, K.H., Drake, A.J., \& Geha, M., 2006,
\apj, 638, 772

\reference{Kallivayalil2}
Kallivayalil, N., van der Marel, R.P., \& Alcock, C., 2006,
\apj, 652, 1213

\reference{Kiziltan}
Kiziltan, B., Baumgardt, H., \& Loeb, A., 2017,
\nat, 542, 203

\reference{Loeb}
Loeb, A., 1998, \apj, 499, L111

\reference{Milone}
Milone, A.P., Marino, A.F., Mastrobuono-Battisti, A., \& Lagioia, E.P., 2018,
\mnras, 479, 5005

\reference{Nusser}
Nusser, A., Banchini, E., Davis, M., 2012, \apj, 755, 58

\reference{Palmer}
Palmer, J.A., Kreutz-Delgado, K., Makeig, S., 2010, Latent
Variable Analysis and Signal Separation, 6365, 303

\reference{Quercellini}
Quercellini, C., Amendola, A., Balbi, A., Cabella, P., Quartin, M., 2012,
Phys. Rep., 521, 95

\reference{Riess}
Riess, A.G., 1999, in eds. Courteau, S.,
Strauss, M.A., Willick, J.A., Cosmic Flows 1999:  Towards an Understanding of Large-Scale
Structure, p. 80

\reference{Sandage}
Sandage, A., 1962, \apj, 136, 319

\reference{Sheth}
Sheth, R.K., Diaferio, A., 2001, \mnras, 322, 901

\reference{Sorce}
Sorce, J.G., 2015, \mnras, 450, 2644

\reference{Springob}
Springob, C.M., et al., 2014, \mnras, 445, 2677

\reference{Tully1}
Tully, R.B., et al., 2013, \apj, 146, 86

\reference{Tully2}
Tully, R.B., Courtois, H.M., \& Sorce, J.G., 2016, \aj, 152, 50

\reference{Watkins}
Watkins, R., \& Feldman, H.A., 2015, \mnras, 450, 1868

\end{references}
\end{document}